# Multi-Cell Processing with Limited Cooperation: A Novel Framework to Timely Designs and Reduced CSI Feedback with General Inputs

Samah A. M. Ghanem, *Member, IEEE*

*Abstract*—We investigate the optimal power allocation and optimal precoding for a multi-cell-processing (MCP) framework with limited cooperation. In particular, we consider two base stations (BSs) which maximize the achievable rate for two users connecting to each BS and sharing channel state information (CSI). We propose a two way channel estimation or prediction process: A pilot assisted estimation at the receiver side, and an autoregressive prediction at the transmitter side. Such framework has promising outcomes in terms of feedback reduction and achievable rates moving the system from one with unknown CSI at the transmitter to a system with instantaneous CSI at both sides of the communication. We derive new extensions of the fundamental relation between the gradient of the mutual information and the MMSE for the conditional and non-conditional mutual information. Capitalizing on such relations, we provide the optimal power allocation and optimal precoding designs with respect to the estimated channel and MMSE. The designs introduced are optimal for multiple access (MAC) Gaussian coherent time-varying fading channels with general inputs and can be specialized to multiple input multiple output (MIMO) channels by decoding interference. The impact of interference on the capacity is quantified by the gradient of the mutual information with respect to the power, channel, and error covariance of the interferer. We provide two novel distributed MCP algorithms that provide the solutions for the optimal power allocation and optimal precoding for the UL and DL with a two way channel estimation to keep track of the channel variations over blocks of data transmission. Therefore, we provide a novel solution that allows with limited cooperation: a significant reduction in the CSI feedback from the receiver to the transmitter, and timely optimal designs of the precoding and power allocation.

## I. Introduction

A system where multiple cooperative base stations jointly serve multiple user terminals (UTs) is referred to as network MIMO system. The MAC channel stands as a special case of interfering channels. In particular, we can model an interference channel with distinct MAC channels, [1]. In the downlink, interference between UTs is already handled at the transmitter side by precoding, where the user data is pre-equalized according to the instantaneous CSI. In [2], the authors exploit a new look into interference via cooperation when precoding is considered. However, the authors in [3] highlighted that the level of interference will decide whether we will decode interference or consider it as noise, which in turn will have direct effect on the optimal power allocation and precoding designs. In [4], the authors show that there are fundamental limits for cooperation, where clusters of limited size should be used. In [5] an ideal limited size MCP sharing both data and CSI has been considered. However, in practice, CSI is typically impaired by channel estimation errors, lossy compression for feedback transmission and feedback delays, [6]. In addition, in practice, cooperative systems incur latency and rate restrictions of backhaul links. In [7], the authors showed that arbitrarily delayed feedback can still allow for performance improvement over the no-CSIT case. Later, the authors in [8], quantified the usefulness of combining delayed and completely obsolete CSIT with immediately available but imperfect CSIT. Further, [9] have derived the degrees-of-freedom (DoF) region, which is tight for a large range of CSIT quality. In [10], the authors proposed a sum rate maximizing precoding solution, which accounts for imperfect CSI shared between cooperative BSs. Further, the authors in [11] derive the optimal power allocation that minimizes the outage probability of block-fading channels with arbitrary inputs. In [12], asymptotic expansions has been utilized to optimize the constrained capacity of multiple-antenna fading coherent channels driven by arbitrary inputs. [13] utilizes lower and upper bounds of the average mutual information of MIMO fading channels driven by arbitrary inputs to design precoding solutions.

In this paper, the framework defined and the solution setup address in a novel way a set of problems which co-exist in a communications framework. First, the paper addresses the problem of limited resources of the backhaul link, therefore full CSI and data sharing is not feasible, and an optimal utilization via limited cooperation, sharing only the CSI between BSs. Second, the paper addresses the problem of huge channel feedback demand from the receiver to the transmitter, which consumes resources and decreases the data rates, whether due to the training phase in the DL, extra introduced interference in the UL, or due to estimation and processing at the receiver side. Therefore, this paper provides a solution which is practically tractable in the setup of time varying fading channels driven by general inputs, and allows the transmitter to assist the receiver in the estimation process of the channel. Third, the paper addresses the problem of sub-optimal selection of the power allocation and precoding strategies due to the time mismatch between the channel feedback from the receiver to the transmitter. Therefore, this work provides a solution which allows timely optimal designs that applies to both the theoretically appealing Gaussian inputs, and the arbitrary inputs which are usually used in practice.

The paper[1] is organized as follows, Section II introduces the system model, the design criterion, and the estimation process for minimal feedback and timely designs. Section III and IV introduce the optimum power allocation and optimal precoding, section V introduces the MCP distributed algorithms with minimal cooperation and channels estimation. Section VI introduces numerical results.

The following notation is employed throughout the paper, boldface uppercase letters denote matrices, lowercase letters denote scalars. The superscript, $(.)^*$, $(.)^T$ and $(.)^\dagger$ denote conjugate, transpose, and conjugate transpose operations. $(.)^\star$ denotes optimum. The operator $\nabla$ denotes the gradient of a scalar function with respect to a variable. $\mathbb{E}[.]$ denotes the expectation operator. $diag(.)$ denotes the diagonal matrix. $\|.\|$ denotes the Euclidean norm of a matrix. Finally, $(.)_{ij}$ denotes the $(i,j)th$ element of a matrix.

## II. SYSTEM MODEL

Consider the scenario where the base stations BS1 and BS2 share CSI of the links of user terminals UT1 and UT2 scheduled to transmit simultaneously and served by each BS, respectively. In a cooperative framework with CSI and data sharing, the BS cooperation can be modeled by a cooperative interference channel or a MIMO channel. However, for a backhaul with finite bandwidth, sharing CSI and data may cause processing overhead on the BSs; therefore, limited cooperation is required sharing only the CSI between the two base stations. To model such limited cooperation framework, we use a divide and concur approach to express such framework. In particular, we divide the interference channel and model it by two MACs, in the uplink and downlink[2], and then select optimal designs. Therefore, in the uplink, BS1 and BS2 will receive from UT1 and UT2 the following receive vectors respectively,

$$\mathbf{y_1} = \sqrt{snr}h_{11}\sqrt{P_1}\mathbf{x_1} + \sqrt{snr}h_{12}\sqrt{P_2}\mathbf{x_2} + \mathbf{n_1} \quad (1)$$

$$\mathbf{y_2} = \sqrt{snr}h_{21}\sqrt{P_1}\mathbf{x_1} + \sqrt{snr}h_{22}\sqrt{P_2}\mathbf{x_2} + \mathbf{n_2} \quad (2)$$

$\mathbf{y_1} \in \mathbb{C}^n$ and $\mathbf{y_2} \in \mathbb{C}^n$ represent the received vectors of complex symbols at BS1 and BS2 respectively, $\mathbf{x_1} \in \mathbb{C}^n$ and $\mathbf{x_2} \in \mathbb{C}^n$ represent the vectors of zero mean uncorrelated complex inputs with $\mathbb{E}\left[\mathbf{x_1}\mathbf{x_1}^\dagger\right] = \mathbb{E}\left[\mathbf{x_2}\mathbf{x_2}^\dagger\right] = \mathbf{I}$, respectively, $\mathbf{n_1} \in \mathbb{C^n}$, and $\mathbf{n_2} \in \mathbb{C^n}$ represent vectors of circularly symmetric complex Gaussian random variables with zero mean and identity covariance, i.e., with $\mathcal{CN}(0, \mathbf{I})$. Additionally, $h_{kl}$ represent the complex coefficients of the Rayleigh fading distributed sub-channels between transmitter $k$ and receiver $l$, where the direct links are with $k = l$, and the interference links are with $k \neq l$; $\sqrt{P_k}$ represent the amplitudes of the transmitted signal from each UT; $snr$ is the gain in the signal to noise ratio due to the channel. Therefore, we can write (1) and (2) as: $\mathbf{y} = \sqrt{\mathbf{snr}}\mathbf{HPx} + \mathbf{n}$, where $(\mathbf{H})_{kl} = h_{kl}$, and $\mathbf{P} = diag(\sqrt{P_1}, \sqrt{P_2})$.

### A. Problem Formulation

Instantaneous capacity is meaningful only when instantaneous CSI is available at the transmitter and the receiver, this takes place either when the channel remains fixed, or when it exhibits slow variations over time, so that it can be considered fixed during a number of transmissions. It is then possible to adapt the transmitter signal to each channel realization to achieve such instantaneous capacity. However, obtaining CSI at the transmitter requires either a feedback channel or the application of the channel reciprocity. The channel estimates at the transmitter may suffer imperfections and the transmitter may also own a delayed version of the CSI.

Therefore, for the sake of problem formulation, we utilize the achievable rate regions of a two user interference channel (IC), [1]. In fact, the upper bounds of the achievable rates for each two-user MAC fading channel decomposing the IC- non cooperative MIMO - are as follows,

$$R_1 \leq \mathbb{E}_{\widehat{\mathbf{H}}} I(\mathbf{x_1}; \mathbf{y_1}|\mathbf{x_2}) \quad (3)$$

$$R_2 \leq \mathbb{E}_{\widehat{\mathbf{H}}} I(\mathbf{x_2}; \mathbf{y_2}|\mathbf{x_1}) \quad (4)$$

$$R_1 + R_2 \leq min\ \left[\mathbb{E}_{\widehat{\mathbf{H}}}I(\mathbf{x_1},\mathbf{x_2};\mathbf{y_1}), \mathbb{E}_{\widehat{\mathbf{H}}}I(\mathbf{x_1},\mathbf{x_2};\mathbf{y_2})\right], \quad (5)$$

Therefore, we consider that BS1 will maximize the mutual information for a constrained capacity, achieved by coding over multiple fading blocks, (for MAC1) as follows:

$$max\ \frac{K - ML}{K}\mathbb{E}_{\widehat{\mathbf{H}}}\left[I(\mathbf{x_1},\mathbf{x_2};\mathbf{y_1}|\widehat{\mathbf{H}})\right] \quad (6)$$

$$Subject\ to,\quad \mathbb{E}_{\widehat{\mathbf{H}}}[P_1(\widehat{\mathbf{H}})] \leq Q_1,\ \mathbb{E}_{\widehat{\mathbf{H}}}[P_2(\widehat{\mathbf{H}})] \leq Q_2,$$
$$P_1(\widehat{\mathbf{H}}) \geq 0\ and\ P_2(\widehat{\mathbf{H}}) \geq 0 \quad (7)$$

And BS2 will maximize the mutual information for a constrained capacity, achieved by coding over multiple fading blocks, (for MAC2) as follows;

$$max\ \frac{K - ML}{K}\mathbb{E}_{\widehat{\mathbf{H}}}\left[I(\mathbf{x_1},\mathbf{x_2};\mathbf{y_2}|\widehat{\mathbf{H}})\right] \quad (8)$$

$$Subject\ to,\quad \mathbb{E}_{\widehat{\mathbf{H}}}[P_1(\widehat{\mathbf{H}})] \leq Q_1,\ \mathbb{E}_{\widehat{\mathbf{H}}}[P_2(\widehat{\mathbf{H}})] \leq Q_2,$$
$$P_1(\widehat{\mathbf{H}}) \geq 0\ and\ P_2(\widehat{\mathbf{H}}) \geq 0 \quad (9)$$

where $K$ is the number of symbols per fading block; $M$ is the number of transmit antennas; $L$ is the number of pilot symbols in the channel estimation process, $P_1$ and $P_2$ are the transmitted power corresponding to each UT, $Q_1$ and $Q_2$ are the average power each UT can use, respectively. We consider point to point channels, therefore, $\widehat{\mathbf{H}}$ is the estimated system channel matrix at a certain time, at the receiver side via pilot-assisted estimation, or at the transmitter side via autoregression, to provide timely designs for the UL and DL[3]. We consider that $M = 1$, and $K >> L = 1$, i.e.,

---

[1]Few results of this paper are presented at IEEE WCNC in [14]

[2]Notice that the MAC in the UL corresponds to two UTs and one BS, and the MAC in the DL corresponds to two BSs and one UT.

[3]Notice that the size of the system channel matrix for the simplest two user case is $2 \times 2$, used for ease of exploitation, however the same setup applies to $k$ users where the system channel matrix scales to $k \times k$

$(K-ML)/K \to 1$ for our analysis. Therefore, the Lagrangian of the optimization problems above are respectively,

$$\mathcal{L}(P_1(\widehat{\mathbf{H}}), P_2(\widehat{\mathbf{H}}), \lambda_1, \lambda_2) = -\mathbb{E}_{\widehat{H}} I(\mathbf{x_1}, \mathbf{x_2}; \mathbf{y_1}|\widehat{\mathbf{H}})$$
$$- \lambda_1(Q_1 - \mathbb{E}_{\widehat{\mathbf{H}}} P_1(\widehat{\mathbf{H}})) - \lambda_2(Q_2 - \mathbb{E}_{\widehat{\mathbf{H}}} P_2(\widehat{\mathbf{H}}))$$
$$- \mu_1 P_1(\widehat{\mathbf{H}}) - \mu_2 P_2(\widehat{\mathbf{H}}) \quad (10)$$

and,

$$\mathcal{L}(P_1(\widehat{\mathbf{H}}), P_2(\widehat{\mathbf{H}}), \lambda_1, \lambda_2) = -\mathbb{E}_{\widehat{H}} I(\mathbf{x_1}, \mathbf{x_2}; \mathbf{y_1}|\widehat{\mathbf{H}})$$
$$- \lambda_1(Q_1 - \mathbb{E}_{\widehat{\mathbf{H}}} P_1(\widehat{\mathbf{H}})) - \lambda_2(Q_2 - \mathbb{E}_{\widehat{\mathbf{H}}} P_2(\widehat{\mathbf{H}}))$$
$$- \mu_1 P_1(\widehat{\mathbf{H}}) - \mu_2 P_2(\widehat{\mathbf{H}}) \quad (11)$$

With primal feasibility conditions, $\lambda_1(Q_1 - \mathbb{E}_{\widehat{\mathbf{H}}} P_1(\widehat{\mathbf{H}})) = 0$, $\mu_1 P_1(\widehat{\mathbf{H}}) = 0$, $\lambda_2(Q_2 - \mathbb{E}_{\widehat{\mathbf{H}}} P_2(\widehat{\mathbf{H}})) = 0$, and $\mu_2 P_2(\widehat{\mathbf{H}}) = 0$. And dual feasibility condition, $\lambda_1 \geq 0$ and $\lambda_2 \geq 0$. Then, applying the KKT conditions we can derive two sets of optimal solutions.

Using the same argument in [1], and capitalizing on the known rate regions of the MAC Gaussian fading channels decomposing the interference fading channel [15], the optimal design selection criterion will be the following,

$$min\left[max\ \mathbb{E}_{\widehat{\mathbf{H}}}\left[I(x_1, x_2; y_1|\widehat{\mathbf{H}})\right], max\ \mathbb{E}_{\widehat{\mathbf{H}}}\left[I(x_1, x_2; y_2|\widehat{\mathbf{H}})\right]\right] \quad (12)$$

Therefore, the optimum power allocation selected set is the solution of the MAC which satisfies a minimum of the maximized mutual information in (6) or (8) subject to the two power constraints (7) and (9).

### B. Two Way Channel Estimation

Obtaining CSI at the transmitter requires either a feedback channel or the application of the channel reciprocity. The channel estimates at the transmitter may suffer imperfections mainly due to a fast time-varying nature of the channel, therefore, the transmitter owns a delayed version of the CSI. Therefore, we consider that channel estimation is done at the receiver side with a pilot assisted mechanism. However, to solve the time mismatch in the design provided at the transmitter, we propose that the transmitter perform estimation of the future channel over the time varying blocks via autoregressive (AR) models, if the coherence time of the channel is large enough and the variation is small, i.e., a slow fading case, we can estimate the future samples over multiple coherence blocks. This proposal will reduce the amount of feedback required from the receiver. However, after the quality of the estimates decreases, the receiver will require to feedback again another pilot to allow the transmitter to predict over more blocks of data transmissions. AR models provide a tool to estimate the dynamics of fading channels, an auto-regressive moving-average (ARMA) model with order $L$, is expressed as,

$$\mathbf{H}(t) = -\sum_{i=1}^{L} \rho \mathbf{H}(t-i) + \mathbf{\Omega}(t), \quad (13)$$

$\mathbf{\Omega} \sim \mathcal{N}(0, \mathbf{I})$ is a zero mean unit variance white Gaussian process, and the AR correlation coefficient bounded as $0 \leq \rho \leq 1$[4]. $T \leq t \leq nT$ corresponds to the time instance at which the channel is estimated or predicted at different periods $n$ of the coherence time $T$. After receiving the pilots, each BS estimates the time varying future channel of the main user and interferer and share the CSI information. Notice that we drop the time index in the rest of the paper.

### III. OPTIMAL POWER ALLOCATION WITH LIMITED COOPERATION

This section presents the characterization of optimal power allocation for the MCP framework with limited cooperation modeled by two distinct MAC Gaussian fading channels and driven by Gaussian, arbitrary, and mixed inputs. The estimation process via auto-regression at the transmitter side, makes possible to estimate the probability of bit error rate in order to minimize it [16]. However, we focus on the maximization of the mutual information as the main design criterion and capitalize on the connections between information measures and estimation measures to devise optimal designs [17], [18], [19].

#### A. Gaussian Inputs

The mutual information for BS1 and BS2, respectively, are defined as,

$$I(\mathbf{x_1}, \mathbf{x_2}; \mathbf{y_1}|\widehat{\mathbf{H}}) = log\left(\frac{|h_{11}|^2 P_1(\widehat{\mathbf{H}}) + |h_{12}|^2 P_2(\widehat{\mathbf{H}})}{\sigma_1^2} + 1\right) \quad (14)$$

where $\sigma_1^2 = 1 + \mathbb{E}[\omega_{11}\omega_{11}^\dagger] + \mathbb{E}[\omega_{12}\omega_{12}^\dagger]$ is the noise covariance introduced due to the prediction process[5].

$$I(\mathbf{x_1}, \mathbf{x_2}; \mathbf{y_2}|\widehat{\mathbf{H}}) = log\left(\frac{|h_{21}|^2 P_1(\widehat{\mathbf{H}}) + |h_{22}|^2 P_2(\widehat{\mathbf{H}})}{\sigma_2^2} + 1\right) \quad (15)$$

where $\sigma_2^2 = 1 + \mathbb{E}[\omega_{21}\omega_{21}^\dagger] + \mathbb{E}[\omega_{22}\omega_{22}^\dagger]$ is the noise covariance introduced due to the prediction process. Therefore, solving the optimization problem for maximizing each MAC achievable rate subject to the users power constraints will lead to the optimal power allocation introduced in the following theorem. Notice that $P_1$, and $P_2$ in the rest of the paper is a function of the channel variation over time, however, we drop this indexing in the rest of the paper.

*Theorem 1:* The optimal power allocation for two UTs in the MCP framework with limited cooperation $(P_1^\star, P_2^\star)$ with Gaussian inputs takes the following form,
For MAC1:

$$\begin{cases} P_1^* = \frac{1}{\lambda_1} - \frac{|h_{12}|^2}{|h_{11}|^2} P_2 - \frac{\sigma_1^2}{|h_{11}|^2}, & \lambda_1 < \frac{|h_{11}|^2}{|h_{12}|^2 P_2 + \sigma_1^2} \\ P_2^* = \frac{1}{\lambda_2} - \frac{|h_{11}|^2}{|h_{12}|^2} P_1 - \frac{\sigma_1^2}{|h_{12}|^2}, & \lambda_2 < \frac{|h_{12}|^2}{|h_{11}|^2 P_1 + \sigma_1^2} \\ P_1^* = 0, & \lambda_1 \geq \frac{|h_{11}|^2}{|h_{12}|^2 P_2 + \sigma_1^2} \\ P_2^* = 0, & \lambda_2 \geq \frac{|h_{12}|^2}{|h_{11}|^2 P_1 + \sigma_1^2} \end{cases}$$
(16)

For MAC2:

---

[4]Note that the pilot sample feedback to the transmitter for the prediction process is assumed to be perfect and noiseless.

[5]Note that the existence of an error in the prediction process introduces a non-Gaussian distributed noise term. However, to make the solution more tractable, we assume that the noise covariance is Gaussian and has negligible effect on the detection process.

$$\begin{cases} P_1^* = \frac{1}{\lambda_1} - \frac{|h_{22}|^2}{|h_{21}|^2}P_1 - \frac{\sigma_2^2}{|h_{21}|^2}, & \lambda_1 < \frac{|h_{21}|^2}{|h_{22}|^2 P_2 + \sigma_2^2} \\ P_2^* = \frac{1}{\lambda_2} - \frac{|h_{21}|^2}{|h_{22}|^2}P_2 - \frac{\sigma_2^2}{|h_{22}|^2}, & \lambda_2 < \frac{|h_{22}|^2}{|h_{21}|^2 P_1 + \sigma_2^2} \\ P_1^* = 0, & \lambda_1 \geq \frac{|h_{21}|^2}{|h_{22}|^2 P_2 + \sigma_2^2} \\ P_2^* = 0, & \lambda_2 \geq \frac{|h_{22}|^2}{|h_{21}|^2 P_1 + \sigma_2^2} \end{cases}$$
(17)

Therefore, $(P_1^\star, P_2^\star)$ is the solution set that satisfies (12), and $\lambda_1^{-1}, \lambda_2^{-1}$ are the water-levels in the waterfilling solution, [20]. Therefore, the numerical solution satisfies the one either for MAC1 or MAC2.

*Proof:* Theorem 1 follows the solution of the Karush-Kuhn-Tucker (KKT) conditions of (6) subject to (7) and (8) subject to (9). ∎

### B. Arbitrary Inputs

To tackle the optimal power allocation and optimal precoding problem with arbitrary inputs, we require the fundamental relation between the mutual information and the minimum mean square error (MMSE), see [18], [19], and [17]. The average system $mmse(snr)$ is given by,

$$\overline{mmse}(snr) = \mathbb{E}_{\widehat{\mathbf{H}}}[mmse(snr, \widehat{\mathbf{H}})] \quad (18)$$
$$= \mathbb{E}_{\widehat{\mathbf{H}}}\left[\mathbb{E}\left[\left\|\widehat{\mathbf{H}}\mathbf{P}(\mathbf{x} - \mathbb{E}[\mathbf{x}|\mathbf{y}])\right\|^2\right]|\widehat{\mathbf{H}}\right]$$

Therefore, we can write the system error matrix $\mathbf{E}$ using the elements of the gradient of the mutual information with respect to the main $m$ and interference links $i$ power. In particular, when each UT uses SISO links, each row of the matrix corresponds to the MMSE of each receiver, therefore, its called, a system MMSE matrix with the elements are given by,

$$\mathbf{E} = \begin{bmatrix} E_{11} & E_{12} \\ E_{21} & E_{22} \end{bmatrix}, \quad (19)$$

with $E_{mm}$ is the error in each direct link, and $E_{mi}$ is the covariance induced due to the interferer link, given by,

$$E_{11} = \mathbb{E}_{\widehat{H}}\left[\mathbb{E}[(\mathbf{x_1} - \mathbb{E}(\mathbf{x_1}|\mathbf{y_1}, \widehat{\mathbf{H}}))(\mathbf{x_1} - \mathbb{E}(\mathbf{x_1}|\mathbf{y_1}, \widehat{\mathbf{H}}))^\dagger]\right] \quad (20)$$

$$E_{12} = \mathbb{E}_{\widehat{H}}\left[\mathbb{E}[(\mathbf{x_1} - \mathbb{E}(\mathbf{x_1}|\mathbf{y_1}, \widehat{\mathbf{H}}))(\mathbf{x_2} - \mathbb{E}(\mathbf{x_2}|\mathbf{y_1}, \widehat{\mathbf{H}}))^\dagger]\right] \quad (21)$$

$$E_{21} = \mathbb{E}_{\widehat{H}}\left[\mathbb{E}[(\mathbf{x_2} - \mathbb{E}(\mathbf{x_2}|\mathbf{y_2}, \widehat{\mathbf{H}}))(\mathbf{x_1} - \mathbb{E}(\mathbf{x_1}|\mathbf{y_2}, \widehat{\mathbf{H}}))^\dagger]\right] \quad (22)$$

$$E_{22} = \mathbb{E}_{\widehat{H}}\left[\mathbb{E}[(\mathbf{x_2} - \mathbb{E}(\mathbf{x_2}|\mathbf{y_2}, \widehat{\mathbf{H}}))(\mathbf{x_2} - \mathbb{E}(\mathbf{x_2}|\mathbf{y_2}, \widehat{\mathbf{H}}))^\dagger]\right] \quad (23)$$

The input estimates are given by,

$$\mathbb{E}(\mathbf{x_k}|\mathbf{y_l}, \widehat{\mathbf{H}}) = \sum_{x_k} \frac{\mathbf{x_k} p_{y_l|x_k, \widehat{H}}(y_l|x_k, \widehat{H}) p(x_k)}{p_{y_l}(\mathbf{y_l})} \quad (24)$$

For the MAC Gaussian time-varying fading channels which corresponds to each MAC created by the limited cooperation framework from each UT to each BS, and driven by arbitrary inputs from each user, we can derive the optimal power allocation for the generalized inputs capitalizing on the relation between the mutual information and the MMSE, [17], [19].

*Theorem 2:* The relation between the gradient of the mutual information in (6) and (8) and with respect to the estimated channel, precoder, and the MMSE matrix for both two user MAC Gaussian fading channel within the interference channel are given by,
For MAC1:

$$\nabla_{P_1} I(\mathbf{x_1}, \mathbf{x_2}; \mathbf{y_1}|\widehat{\mathbf{H}}) = |\widehat{h}_{11}|^2 \sqrt{P_1} E_{11} + \widehat{h}_{11}^* \widehat{h}_{12} \sqrt{P_2} E_{12} \quad (25)$$

$$\nabla_{P_2} I(\mathbf{x_1}, \mathbf{x_2}; \mathbf{y_1}|\widehat{\mathbf{H}}) = |\widehat{h}_{12}|^2 \sqrt{P_2} E_{12} + \widehat{h}_{12}^* \widehat{h}_{11} \sqrt{P_1} E_{11} \quad (26)$$

For MAC2:

$$\nabla_{P_1} I(\mathbf{x_1}, \mathbf{x_2}; \mathbf{y_2}|\widehat{\mathbf{H}}) = |\widehat{h}_{21}|^2 \sqrt{P_1} E_{12} + \widehat{h}_{21}^* \widehat{h}_{22} \sqrt{P_2} E_{22} \quad (27)$$

$$\nabla_{P_2} I(\mathbf{x_1}, \mathbf{x_2}; \mathbf{y_2}|\widehat{\mathbf{H}}) = |\widehat{h}_{22}|^2 \sqrt{P_2} E_{22} + \widehat{h}_{22}^* \widehat{h}_{21} \sqrt{P_1} E_{21} \quad (28)$$

*Proof:* Theorem 2 is a direct consequence to Theorem 6, in [17]. ∎

Theorem 2 shows how much rate is lost due to interference links, this is due to the fact that some terms in the gradient of the mutual information preclude the effect of the mutual interference of the direct links,see [5]. Therefore, we can account for such quantified rate loss via optimal power allocation. Then, in order to be able to attack the optimization problem we capitalize on the chain rule of the mutual information to derive the conditional mutual information as follows,

$$I(\mathbf{x_1}, \mathbf{x_2}; \mathbf{y_1}|\widehat{\mathbf{H}}) = I(\mathbf{x_2}; \mathbf{y_1}|\widehat{\mathbf{H}}) + I(\mathbf{x_1}; \mathbf{y_1}|\mathbf{x_2}, \widehat{\mathbf{H}}) \quad (29)$$

and,

$$I(\mathbf{x_1}, \mathbf{x_2}; \mathbf{y_2}|\widehat{\mathbf{H}}) = I(\mathbf{x_1}; \mathbf{y_2}|\widehat{\mathbf{H}}) + I(\mathbf{x_2}; \mathbf{y_2}|\mathbf{x_1}, \widehat{\mathbf{H}}) \quad (30)$$

Clearly, we know that $I(\mathbf{x_2}; \mathbf{y_1}|\widehat{\mathbf{H}})$ is the mutual information when UT2 is decoded considering UT1 signal as noise. Therefore, we can write it as follows,

$$I(\mathbf{x_2}; \mathbf{y_1}|\widehat{\mathbf{H}}) = \mathbb{E}\left[\log \frac{p_{y_1|x_2, \widehat{H}}(\mathbf{y_1}|\mathbf{x_2}, \widehat{\mathbf{H}})}{p_{y_1}(\mathbf{y_1})}\right] \quad (31)$$

$$p_{y_1|x_2, \widehat{H}}(\mathbf{y_1}|\mathbf{x_2}, \widehat{\mathbf{H}}) = \frac{1}{|\widehat{h}_{11}|^2 P_1 + 1} e^{-\frac{\left\|\mathbf{y_1} - \sqrt{snr}\widehat{h}_{12}\sqrt{P_2}\mathbf{x_2}\right\|^2}{2(|\widehat{h}_{11}|^2 P_1 + 1)}} \quad (32)$$

$$p_{y_1}(\mathbf{y_1}) = \sum_{x_2} p_{y_1|x_2, \widehat{H}}(\mathbf{y_1}|\mathbf{x_2}, \widehat{\mathbf{H}}) p_{x_2}(\mathbf{x_2}) \quad (33)$$

Based on such definition, we conclude the following theorem which provides a new fundamental relation between the mutual information of the arbitrary distributed interfering signal and the direct link received vector.

*Theorem 3:* The gradient of the mutual information with respect to the power allocation, for a scaled user power by considering the interference as noise is as follows,

$$\nabla_{P_1} I(\mathbf{x_2}; \mathbf{y_1}|\widehat{\mathbf{H}}) = \frac{1}{(|\widehat{h}_{11}|^2 P_1 + 1)} \widehat{h}_{11}^2 \widehat{h}_{12}^2 P_2 E_{22} \quad (34)$$

$$\nabla_{P_2} I(\mathbf{x_1}; \mathbf{y_2}|\widehat{\mathbf{H}}) = \frac{1}{(|\widehat{h}_{22}|^2 P_2 + 1)} \widehat{h}_{22}^2 \widehat{h}_{21}^2 P_1 E_{11} \quad (35)$$

*Proof:* See Appendix A. ∎

We can similarly derive the gradient with respect to the power allocation $\nabla_{P_1} I(\mathbf{x_1}; \mathbf{y_2}|\widehat{\mathbf{H}})$, $\nabla_{P_1} I(\mathbf{x_1}; \mathbf{y_2}|\widehat{\mathbf{H}})$, $\nabla_{P_2} I(\mathbf{x_2}; \mathbf{y_2}|\widehat{\mathbf{H}})$ and $\nabla_{P_2} I(\mathbf{x_2}; \mathbf{y_2}|\widehat{\mathbf{H}})$. Additionally, in order to find the gradient of the conditional mutual information, an easier approach is to capitalize on the chain rule of the mutual information, as concluded in the following corollary.

*Corollary 1:* The gradient of the conditional mutual information will be as follows,

$$\nabla_{P_1} I(\mathbf{x_1}; \mathbf{y_1}|\mathbf{x_2}, \widehat{\mathbf{H}}) = \nabla_{P_1} I(\mathbf{x_1}, \mathbf{x_2}; \mathbf{y_1}|\widehat{\mathbf{H}}) - \nabla_{P_1} I(\mathbf{x_2}; \mathbf{y_1}|\widehat{\mathbf{H}}) \quad (36)$$

$$\nabla_{P_2} I(\mathbf{x_2}; \mathbf{y_2}|\mathbf{x_1}, \widehat{\mathbf{H}}) = \nabla_{P_2} I(\mathbf{x_1}, \mathbf{x_2}; \mathbf{y_2}|\widehat{\mathbf{H}}) - \nabla_{P_2} I(\mathbf{x_1}; \mathbf{y_2}|\widehat{\mathbf{H}}) \quad (37)$$

*Proof:* The proof of the corollary follows from the chain rule of mutual information and the gradient of the mutual information derived for the sum rate and non-conditional rates. The following theorem derives the optimal power allocation that maximizes both sum rates of each MAC within the interference channel. ∎

*Theorem 4:* The optimal power allocation for both two user MAC Gaussian time-varying fading channel within the interference fading channel and driven by arbitrary inputs - in terms of estimated channels, the interferer power, and the MMSE matrix - takes the following form,
For MAC1:

$$\sqrt{P_1^\star} = \lambda_1^{-1}|\widehat{h}_{11}|^2\sqrt{P_1^\star}E_{11} + \lambda_1^{-1}\widehat{h}_{11}^*\widehat{h}_{12}\sqrt{P_2^\star}E_{12} \quad (38)$$

$$\sqrt{P_2^\star} = \lambda_2^{-1}|\widehat{h}_{12}|^2\sqrt{P_2^\star}E_{12} + \lambda_2^{-1}\widehat{h}_{12}^*\widehat{h}_{11}\sqrt{P_1^\star}E_{11} \quad (39)$$

For MAC2:

$$\sqrt{P_1^\star} = \lambda_1^{\star-1}|\widehat{h}_{21}|^2\sqrt{P_1^\star}E_{21} + \lambda_1^{\star-1}\widehat{h}_{21}^*\widehat{h}_{22}\sqrt{P_2^\star}E_{22} \quad (40)$$

$$\sqrt{P_2^\star} = \lambda_2^{\star-1}|\widehat{h}_{22}|^2\sqrt{P_2^\star}E_{22} + \lambda_2^{\star-1}\widehat{h}_{22}^*\widehat{h}_{21}\sqrt{P_1^\star}E_{21} \quad (41)$$

Therefore, $(P_1^\star, P_2^\star)$ is the solution set that satisfies (12), and $\lambda_1, \lambda_2$ are Lagrange multipliers normalized by $snr$. Therefore, the numerical solution satisfies the one either for MAC1 or MAC2.

*Proof:* The proof of Theorem 4 follows the KKT conditions solving (6) subject to (7) and (8) subject to (9). ∎

Its straightforward to see that, $P_2^\star = Q_2$, when $P_1 = 0$, and $P_1^\star = Q_1$ when $P_2 = 0$. In addition, we can specialize the result of Theorem 3 to the one for Gaussian inputs. In particular, we substitute the elements of the linear MMSE for Gaussian inputs into (38) and (39), then the optimal power allocation in Theorem 3 matches the one in Theorem 1. Theorem 4 assimilates a mercury/waterfilling for the arbitrary inputs that compensate for the non-Gaussianess of the binary constellations, and a waterfilling for the Gaussian inputs. Moreover, we can re-write Theorem 3 with respect to the MMSE and the covariance, for the main and interference of each SISO MAC as follows,

$$P_m^\star = \frac{1}{snr|\widehat{h}_m|^2} mmse_m(snr|\widehat{h}_m|^2 P_m^\star)$$
$$+ \frac{1}{snr\widehat{h}_m^*\widehat{h}_i} cov(snr|\widehat{h}_m^*\widehat{h}_i|P_i^\star) \quad (42)$$

$$P_i^\star = \frac{1}{snr|\widehat{h}_i^*\widehat{h}_m|} mmse_m(snr|\widehat{h}_i^*\widehat{h}_m|P_m^\star)$$
$$+ \frac{1}{snr|\widehat{h}_i|^2} cov(snr|\widehat{h}_i|^2 P_i^\star), \quad (43)$$

### C. Mixed Inputs

The mixed inputs case is the case where one input is Gaussian and the other is BPSK, this assumption is relevant for instance, when an unfriendly jammer is trying to inject interference in the form of Gaussian or arbitrary noise, see [21]. We need to capitalize on the previous derived tools to address the solution of this setup. The implementation of such scenario can be tackled via the chain rule of the mutual information as follows,

$$I(\mathbf{x_1}, \mathbf{x_2}; \mathbf{y_1}|\widehat{\mathbf{H}}) = \mathbf{I}(\mathbf{x_1}; \mathbf{y_1}|\widehat{\mathbf{H}}) + \mathbf{I}(\mathbf{x_2}; \mathbf{y_1}|\mathbf{x_1}, \widehat{\mathbf{H}}) \quad (44)$$

If $\mathbf{x_1}$ corresponds to the BPSK input, and $\mathbf{x_2}$ corresponds to the Gaussian input, then we can find the achievable rate of BS1; assuming that BS1 satisfies (12), then we can re-write (44) as follows,

$$I(\mathbf{x_1}, \mathbf{x_2}; \mathbf{y_1}|\widehat{\mathbf{H}}) = I(\mathbf{x_1}; \mathbf{y_1}|\widehat{\mathbf{H}}) + log(|\widehat{h}_{12}|^2 P_2 + 1), \quad (45)$$

with,
$$I(\mathbf{x_1}; \mathbf{y_1}|\widehat{\mathbf{H}}) = \mathbb{E}\left[\log \frac{p_{\mathbf{y_1}|\mathbf{x_1}, \widehat{H}}(\mathbf{y_1}|\mathbf{x_1}, \widehat{\mathbf{H}})}{p_{\mathbf{y_1}}(\mathbf{y_1})}\right] \quad (46)$$

$$p_{y_1|x_1,\widehat{H}}(\mathbf{y_1}|\mathbf{x_1}, \widehat{\mathbf{H}}) = \frac{1}{|\widehat{h}_{12}|^2 P_2 + 1} e^{-\frac{\|\mathbf{y_1} - \sqrt{snr}\widehat{h}_{11}\sqrt{P_1}\mathbf{x_1}\|^2}{2(|\widehat{h}_{12}|^2 P_2 + 1)}} \quad (47)$$

$$p_{y_1}(\mathbf{y_1}) = \sum_{x_1} p_{y_1|x_1,\widehat{H}}(\mathbf{y_1}|\mathbf{x_1}, \widehat{\mathbf{H}}) p_{x_1}(\mathbf{x_1}) \quad (48)$$

Its not anymore straightforward to use the relation between the mutual information and the MMSE for such setup of mixed inputs. However, we need to consider two components of the derivative with respect to the SNR: The derivative of the conditional mutual information corresponding to the Gaussian input equals the linear MMSE, i.e., $mmse_2(snr)$. The other part of the derivative corresponds to the arbitrary input with an SNR scaled with the other input channel and power. The implication is a scaled MMSE for user 1, which is in turn, in the form of $mmse_1(snr\frac{|\widehat{h}_{11}|^2 P_1}{|\widehat{h}_{12}|^2 P_2 + 1})$. We can easily deduce that the optimal power allocation for the mixed input case is a mixture of the two policies presented for the Gaussian and arbitrary inputs. Further details on the generalization of the known fundamental relation between the mutual information and the MMSE to the multiuser setup can be found in [22], [23] and [24].

## IV. OPTIMAL PRECODING WITH LIMITED COOPERATION

Consider the MCP with limited cooperation in the DL where both BSs cooperate sharing their CSI estimated versions to design the optimal precoding vectors that maximize the system achievable rate. The following theorem gives a generalized optimal precoder for the MCP with limited cooperation. In particular, this theorem provides an optimal precoding set for BS1 and BS2, which can be generalized to setups with multiple cooperating base stations and multiple user MAC channels.

*Theorem 5:* The non-unique optimal precoding set that maximizes the achievable rates for both two user MAC Gaussian fading channel within the interference channel subject to per user power constraint is the numerical solution of the following form,
For MAC1:

$$\mathbf{P_1}^\star = \nu_1^{-1}\widehat{\mathbf{H}}_{11}^\dagger \widehat{\mathbf{H}}_{11}\mathbf{P_1}^\star \mathbf{E_{11}} + \nu_1^{-1}\widehat{\mathbf{H}}_{11}^\dagger \widehat{\mathbf{H}}_{12}\mathbf{P_2}^\star \mathbf{E_{12}} \quad (49)$$

$$\mathbf{P_2}^\star = \nu_2^{-1}\widehat{\mathbf{H}}_{12}^\dagger \widehat{\mathbf{H}}_{12}\mathbf{P_2}^\star \mathbf{E_{12}} + \nu_2^{-1}\widehat{\mathbf{H}}_{12}^\dagger \widehat{\mathbf{H}}_{11}\mathbf{P_1}^\star \mathbf{E_{11}} \quad (50)$$

For MAC2:

$$\mathbf{P_1}^\star = \nu_1^{-1}\widehat{\mathbf{H}}_{21}^\dagger\widehat{\mathbf{H}}_{21}\mathbf{P_1}^\star\mathbf{E_{21}} + \nu_1^{-1}\widehat{\mathbf{H}}_{21}^\dagger\widehat{\mathbf{H}}_{22}\mathbf{P_2}^\star\mathbf{E_{22}} \quad (51)$$

$$\mathbf{P_2}^\star = \nu_2^{-1}\widehat{\mathbf{H}}_{22}^\dagger\widehat{\mathbf{H}}_{22}\mathbf{P_2}^\star\mathbf{E_{22}} + \nu_2^{-1}\widehat{\mathbf{H}}_{22}^\dagger\widehat{\mathbf{H}}_{21}\mathbf{P_1}^\star\mathbf{E_{21}} \quad (52)$$

Therefore, $(\mathbf{P_1}^\star, \mathbf{P_2}^\star)$ is the solution set that satisfies (12), and $\nu_1, \nu_2$ are the per MAC *snr* normalized by the Lagrange multipliers.

*Proof:* Theorem 5 follows the relation between the gradient of the mutual information and the MMSE and the decomposition of its matrix components[6]. ∎

In this limited cooperation scenario, its worth to consider that each BS try to decode the interference rather than sharing the global data with the other BS. Thus, each input can be estimated at the processing BS/UT. Moreover, we can preserve the optimal properties of the iterative solution in Theorem 5 and through decoding interference, we can preclude the effect of the covariance terms. Therefore, the optimal precoder takes the fixed point equation optimal solution of MIMO channels $\mathbf{P}^\star = \nu^{-1}\widehat{\mathbf{H}}^\dagger\widehat{\mathbf{H}}\mathbf{P}^\star\mathbf{E}$, see [5], [25], [26], [27].

Its worth to notice that the second covariance term in (38) and (39) quantifies the power cost we need to pay due to interference. However, the second term in equations (25) to (26) in the gradient quantifies the losses in terms of information rates due to interference.

## V. MCP WITH LIMITED COOPERATION DISTRIBUTED ALGORITHMS

We introduce the MCP with limited cooperation distributed algorithms, the first algorithm gives the optimal power allocation for the UL, and the second algorithm gives the optimal precoding for the DL.

## VI. NUMERICAL ANALYSIS

We shall now introduce a set of illustrative results that cast some insights into the problem. We analyze the results for the system with cooperation limited MACs under three different types of inputs and under a fixed and instantaneous estimated channel. We focus our analysis on real channels to establish certain facts about the rate losses due to interference. The results for the Gaussian inputs setup are straightforward with the mutual information closed form. However, we used Monte-Carlo method to generate the achievable rates for arbitrary inputs and mixed inputs.

Figure 1, Figure 2, and Figure 3 show the achievable rates for one BS in the MAC setup -supposing that this achievable rate is the one that provides the minimum of the maximum of the achievable rates of both BSs-using Gaussian inputs, BPSK inputs, and mixed inputs. In the case where both inputs are arbitrary, for example; BPSK, the achievable rate faces decay at equal input powers when both user channels are equal and

---

**Algorithm 1:** Optimum Power Allocation with MCP-UL Limited cooperation: CSI sharing

**BS1 Input:** CSI1: $\hat{h}_{11}(t=1),...,\hat{h}_{11}(t=L)$, $\hat{h}_{21}(t=1),...,\hat{h}_{21}(t=L)$, $\mathbb{E}[x_1|y_1]$

**BS2 Input:** CSI2: $\hat{h}_{22}(t=1),...,\hat{h}_{22}(t=L)$, $\hat{h}_{12}(t=1),...,\hat{h}_{12}(t=L)$, $\mathbb{E}[x_2|y_2]$

**if** *BW Backhaul* $\geq Threshold\ \tau$ **then**
| BS1 and BS2 declare congestion and no cooperation
**else**
| BS1 sends CSI1 to BS2, BS2 sends CSI2 to BS1

**Output**:
BS1 and BS2 find the optimum power allocation $(P_1^\star, P_2^\star)$ in the UL as the solution for:

$$\mathbf{P_{m_{k+1}}} = \alpha_k \mathbf{P_{m_k}} + \alpha_k \lambda \widehat{\mathbf{H}}_m^\dagger \widehat{\mathbf{H}}_m \mathbf{P_{m_k}} \mathbf{E_{m_k}} + \alpha_k \lambda \widehat{\mathbf{H}}_m^\dagger \widehat{\mathbf{H}}_i \mathbf{P_{i_k}} \mathbf{E_{i_k}}$$

BS1 and BS2 check resources → handshaking → BS1 and BS2 jointly decide
$(P_1^\star, P_2^\star)$ that satisfies (12)
BS1 and BS2 feedback $P_1^\star$ and $P_2^\star$ to UT1 and UT2, respectively.

---

**Algorithm 2:** Optimum Precoding with MCP-DL Limited cooperation: CSI sharing

**BS1 Input:** CSI1, $x_1$
**BS2 Input:** CSI2, $x_2$

**AR function:** BS1 and BS2 estimates channel variation for the block length $K$ using autoregression of order $L$, $AR_L(ML,...,K-ML)$

**SVD function:** BS1 and BS2 performs
SVD($\widehat{\mathbf{H}}$):$\widehat{\mathbf{H}} = \mathbf{U}_{\hat{H}}\Lambda_{\hat{H}}\mathbf{V}_{\hat{H}}^\dagger$

BS1 sends CSI1 to BS2 and BS2 sends CSI2 to BS1

**Output**:
BS1 and BS2 find the optimum precoding $(P_1^\star, P_2^\star)$ in the DL as the solution for:
$$\mathbf{P_{m_k}} = \mathbf{V}_{\hat{\mathbf{H}}_m}\mathbf{P_m}^{1/2}$$
$$\mathbf{P_{m_{k+1}}} = \alpha_k \mathbf{P_{m_k}} + \alpha_k \lambda \widehat{\mathbf{H}}_m^\dagger \widehat{\mathbf{H}}_m \mathbf{P_{m_k}} \mathbf{E_{m_k}}$$

BS1 and BS2 select jointly the optimal set that satisfies (12),
**BS1 transmits** : $(h_{11}\nu_{\hat{h}_{11}}\sqrt{P_1} + h_{12}\nu_{\hat{h}_{21}}\sqrt{P_1})x_1$
**BS2 transmits** : $(h_{21}\nu_{\hat{h}_{12}}\sqrt{P_2} + h_{22}\nu_{\hat{h}_{22}}\sqrt{P_2})x_2$

The process will be iteratively repeated for each simultaneous transmission of BS1 and BS2.

$UT1$ and $UT2$ receives the block of $K$ symbols, estimate $L$ pilots and feedback to BS1 and BS2

---

real; for instance, if $h_{11} = h_{12} = h_{21} = h_{22} = 1$[7], and at

---

[6]Notice that we abuse the notation, since we provide the general solution in Theorem 5 using full channel matrices, while the model presents scalar channels. This is to show that the derivations apply to the case when the UTs use more than one channel for transmission.

[7]The selection of such values for our preliminary analysis is to allow for symmetric links which makes the rate regions of both MAC channels enclosing the interference channel to coincide.

equal power, both user inputs stay in the null space of the channel; therefore, they encounter a rate loss of 0.5 bits. One way to overcome this is by orthogonalizing the inputs in the UL (and/or precoding in the DL). However it is clear that the achievable rates when inputs are mixed are below those when inputs are Gaussian. This is due to the increase in the scaled MMSE of each input. The optimal power allocation is not presented here due to space limits, however, its clear from the mutual information results with respect to the power of the two UTs that:

In the Gaussian setup, the power allocation is to select the maximum power for this instantaneous estimated/predicted channel, this is due to strong interference at equal channel gains.

In the arbitrary setup using BPSK inputs, it is clear that orthogonalizing inputs, or an unbalanced power allocation allow both inputs not to stay in the null space of the channel, see also [5]. Hence, under strong interference, one input will select to deviate from the maximum power selection, and so the decay in the mutual information at the equal power set, e.g. (2,2), at $45^o$ line shown can be improved.

In the mixed input case, the optimal power allocation is similar to the Gaussian case, where users allocate their maximum power under strong interference.

Figure 4 and Figure 5 show the implication of the decoding sequence on the achievable rates, it is clear that the user distribution firstly decoded has the dominant effect. The rate region shifts towards the user decoded last and so if $x_1$ is first decoded given that $x_2$ is noise, then the rates achieved by user 1 will be as shown in Figure 6. Further, its clear that an unbalanced power allocation (mercury waterfilling) allows for higher achievable rates, especially under such instantaneous knowledge/prediction at the transmitter. More specifically, user 2 decoded last enjoys more rate, almost surely, as far as the powers used by both users are not equal they stay achieving 2 $bits/sec/Hz$, however, as much the power goes to be equal as Figure 4 depicts, the sum rate will decay due to the covariace caused from the users or the called interference to 1.5 $bits/sec/Hz$. Similar results can be obtained from Figure 3. This explains well why under a MCP with joint processing the rates are better as Figure 2 illustrates.

Its worth to notice that it has been recently shown that the user decoded given the knowledge of the other, or technically the user with the conditional term of the mutual information will always enjoy better rates, however, the losses will be incured mainly by the user decoded first. A closed form expressions of the mutual information for conditional and non-conditional users in such setup with BPSK inputs has been provided in [23].

Further, to complete this analysis, we can also see verify from Figure 7 that illustrates the achievable rate of the Gaussian distributed user if decoded first, under the mixed inputs setup, where sum rates are achievable, i.e., the sum rates shown in Figure 3 are not anymore achievable, i.e., if from a theoretical point of view, a Gaussian distributed input exists in conjunction with an arbitrary distributed interferer, its better to decode first the arbitrary input then last the Gaussian input. This explain well also that an arbitrary distributed interference is more harmful than the Gaussian one.

The average and instantaneous mutual information as well as the average and instantaneous MMSE are presented in Figure 8, and Figure 9. We average over 250 channel realizations obtained via first order autoregression with $\rho = 1$; its of particular relevance to see; First, that its better to have less but not least number of realizations to provide a better estimate of the instantaneous channel. Second, we can analyze other aspects related to the fact that diagonalizing the interference channel and so the system error matrix is not an optimal solution for binary constellations. Third, through the estimation of the instantaneous channel its possible not only to provide timely designs, but also to quantify the losses incurred in the data rates, therefore we may avoid going into outage if we have good channel estimates at both sides of the communication.

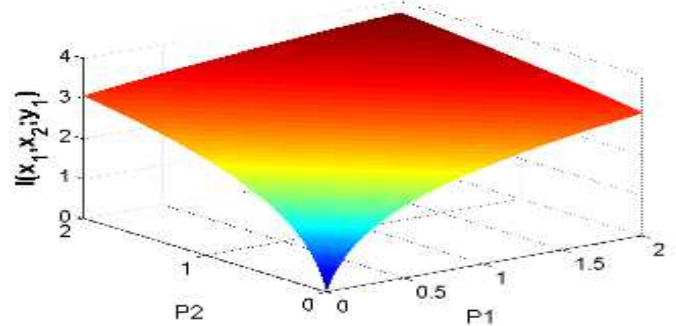

Figure 1. The achievable rate for MAC1 with respect to UTs main power-Gaussian inputs.

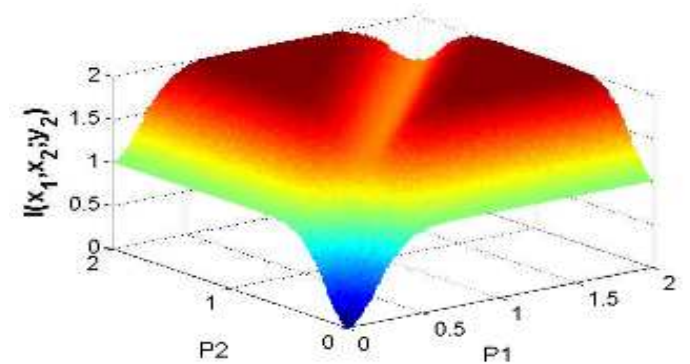

Figure 2. The achievable rate for MAC1 with respect to UTs main power-BPSK inputs.

## VII. CONCLUSION

We have addressed the problem of optimal power allocation and optimal precoding in a limited cooperation framework. We propose a two way channels estimation process that allows the transmitter to foresee the channel variation over the blocks of transmission via a prior-knowledge of the channel distribution and a pilot-assisted channel estimation at the receiver side, aiming to minimize the imperfections of the CSI at the transmitter which cause untimely designs. Therefore, we also allow a reduced feedback overhead which can be of practical relevance to systems with large number of antennas, e.g., like massive MIMOs. Therefore, the proposed solutions indeed break down into the ones with instantaneous knowledge of the channel, allowing limited cooperation, reduced feedback, and timely optimal designs.

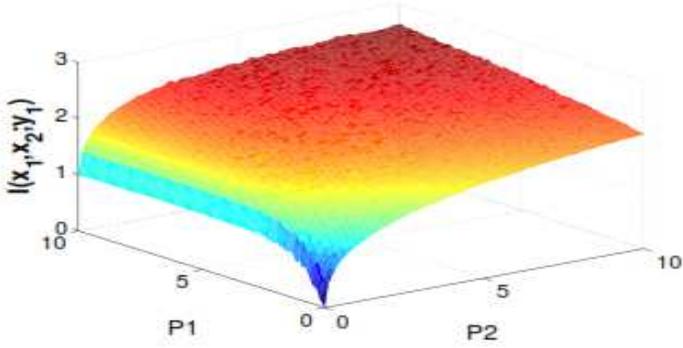

Figure 3. The achievable rate for MAC1 with respect to UTs main power- Mixed inputs, BPSK input $x_1$ is decoded first.

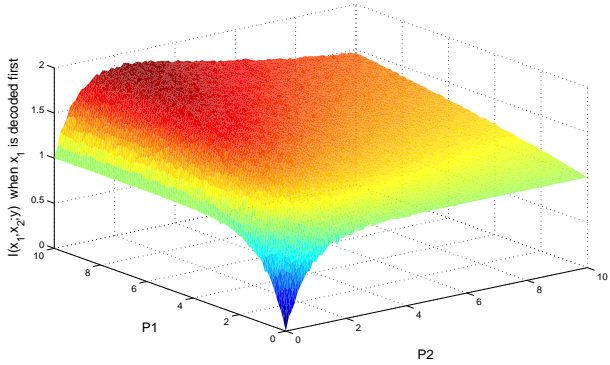

Figure 4. The achievable rate for MAC1 with respect to UTs main power- BPSK inputs, $x_1$ is decoded first.

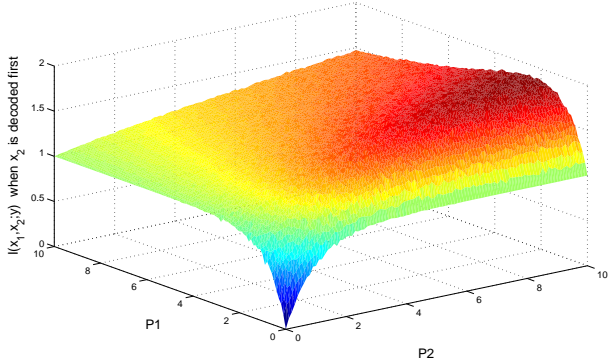

Figure 5. The achievable rate for MAC1 with respect to UTs main power- BPSK inputs, $x_2$ is decoded first.

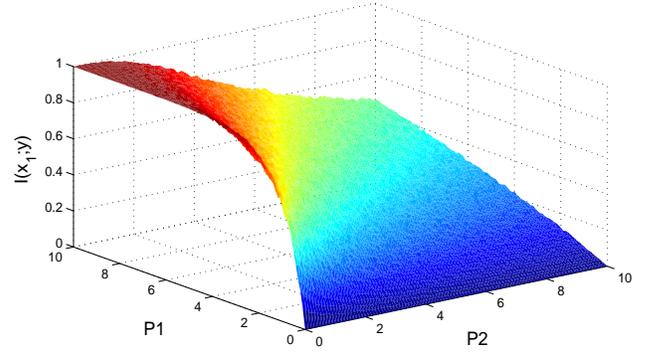

Figure 6. The achievable rate for BPSK $x_1$ of UT1 decoded first with respect to UTs main power. $x_2$ decoded last

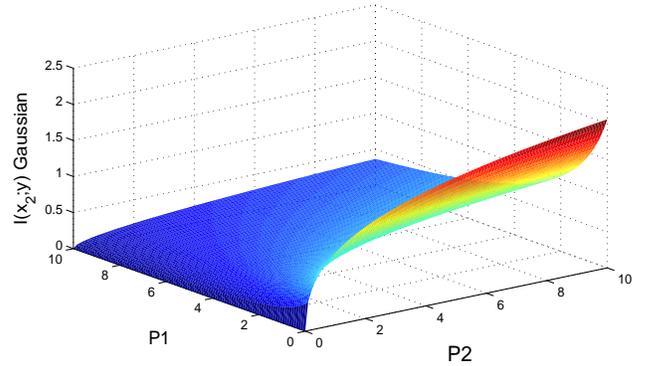

Figure 7. The achievable rate for Gaussian $x_2$ of UT2 decoded first with respect to UTs main power. $x_1$ decoded last

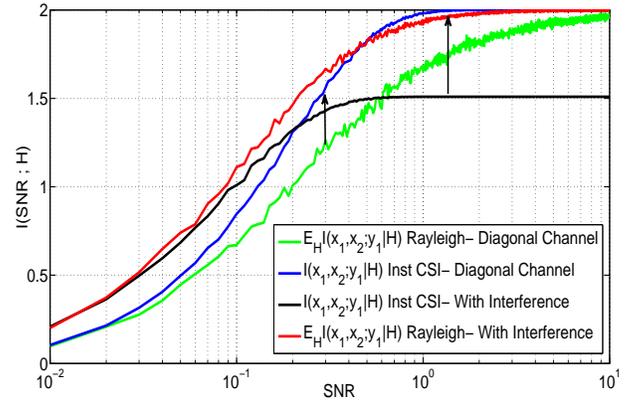

Figure 8. Average and instantaneous mutual information for BPSK inputs under Rayleigh fading and real (for diagonal and interference channels), with $Q1 = Q2 = 2$.

## VIII. ACKNOWLEDGMENT

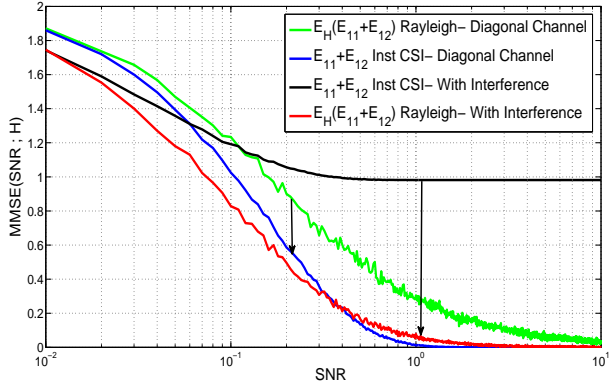

Figure 9. Average and instantaneous MMSE for BPSK inputs under Rayleigh fading and real (for diagonal and interference channels), with $Q1 = Q2 = 2$.

## A. Appendix A: Proof of Theorem 3

The gradient of the mutual information with respect to the power allocation of user 2 on the two-user MAC given that user 1 is noise is as follows:

$$\frac{\partial I(\mathbf{x_2}; \mathbf{y})}{\partial P_1} = -\frac{\partial}{\partial P_1} \int p_{\mathbf{y_1}}(\mathbf{y_1}) log\left(p_{y_1}(\mathbf{y_1})\right) d\mathbf{y_1} \quad (53)$$

$$= -\int \left(p_{y_1}(\mathbf{y_1}) \frac{1}{p_{y_1}(\mathbf{y_1})} + log\left(p_{y_1}(\mathbf{y_1})\right)\right) \frac{\partial p_{y_1}(\mathbf{y_1})}{\partial P_1} d\mathbf{y_1} \quad (54)$$

$$= -\int \left(1 + log\left(p_{y_1}(\mathbf{y_1})\right)\right) \frac{\partial p_{y_1}(\mathbf{y_1})}{\partial P_1} d\mathbf{y_1} \quad (55)$$

The derivative of the conditional output can be written as:

$$\frac{\partial p_{y_1|x_2}(\mathbf{y_1}|\mathbf{x_2})}{\partial P_1} =$$
$$p_{y_1|x_2}(\mathbf{y_1}|\mathbf{x_2}) \frac{\partial}{\partial P_1} \frac{\left(\mathbf{y_1} - \sqrt{snr}h_{12}\sqrt{P_2}\mathbf{x_2}\right)^2}{h_{11}^2 P_1 + 1} \quad (56)$$

$$= -p_{y_1|x_2}(\mathbf{y_1}|\mathbf{x_2}) \frac{h_{11}^2 \left(\mathbf{y_1} - \sqrt{snr}h_{12}\sqrt{P_2}\mathbf{x_2}\right)^2}{(h_{11}^2 P_1 + 1)^2} \quad (57)$$

$$= -\nabla_{\mathbf{y_1}} p_{y_1|x_2}(\mathbf{y_1}|\mathbf{x_2}) \frac{h_{11}^2 \left(\mathbf{y_1} - \sqrt{snr}h_{12}\sqrt{P_2}\mathbf{x_2}\right)}{(h_{11}^2 P_1 + 1)} \quad (58)$$

Therefore, we have:

$$\mathbb{E}_{x_2}\left[\nabla_{P_1} p_{y_1|x_2}(\mathbf{y_1}|\mathbf{x_2})\right] =$$
$$\mathbb{E}_{x_2}\left[\nabla_{\mathbf{y_1}} p_{y_1|x_2}(\mathbf{y_1}|\mathbf{x_2}) \frac{h_{11}^2 \left(\sqrt{snr}h_{12}\sqrt{P_2}\mathbf{x_2}\right)}{(h_{11}^2 P_1 + 1)}\right] \quad (59)$$

Substitute (59) into (55), we get:

$$\frac{\partial I(\mathbf{x_2}; \mathbf{y_1})}{\partial P_1} = \frac{\sqrt{snr}h_{11}^2}{(h_{11}^2 P_1 + 1)} \int (1 + log\left(p_{y_1}(\mathbf{y_1})\right)) \times$$
$$\mathbb{E}_{x_2}\left[\nabla_{\mathbf{y_1}} p_{y_1|,x_2}(\mathbf{y_1}|\mathbf{x_2}) h_{12}\sqrt{P_2}\mathbf{x_2}\right] d\mathbf{y_1} \quad (60)$$

Using integration by parts applied to the real and imaginary parts of $\mathbf{y_1}$, and due to the fact that $||\mathbf{y_1}|| \to \infty$, we have:

$$\frac{\partial I(\mathbf{x_2}; \mathbf{y_1})}{\partial P_1} = \frac{\sqrt{snr}h_{11}^2}{(h_{11}^2 P_1 + 1)} \times$$
$$\mathbb{E}_{x_2}\left[\int (1+log(p_{y_1}(\mathbf{y_1})))\nabla_{\mathbf{y_1}} p_{y_1|x_2}(\mathbf{y_1}|\mathbf{x_2}) d\mathbf{y_1} h_{12}\sqrt{P_2}\mathbf{x_2}\right] \quad (61)$$

Therefore,

$$\frac{\partial I(\mathbf{x_2}; \mathbf{y_1})}{\partial P_1} = \frac{\sqrt{snr}h_{11}^2}{(h_{11}^2 P_1 + 1)} \times$$
$$\mathbb{E}_{x_2}\left[-\int \left(\frac{p_{y_1|x_2}(\mathbf{y_1}|\mathbf{x_2})}{p_{y_1}(\mathbf{y_1})} \nabla_{\mathbf{y_1}} p_{y_1}(\mathbf{y_1}) d\mathbf{y_1}\right) h_{12}\sqrt{P_2}\mathbf{x_2}\right] \quad (62)$$

$$\frac{\partial I(\mathbf{x_2}; \mathbf{y_1})}{\partial P_1} = -\frac{\sqrt{snr}h_{11}^2}{(h_{11}^2 P_1 + 1)} \int \nabla_{\mathbf{y_1}} p_{y_1}(\mathbf{y_1}) \times$$
$$\mathbb{E}_{x_2}\left[\frac{p_{y_1|x_2}(\mathbf{y_1}|\mathbf{x_2})}{p_{y_1}(\mathbf{y_1})} h_{12}\sqrt{P_2}\mathbf{x_2}\right] d\mathbf{y_1} \quad (63)$$

$$\frac{\partial I(\mathbf{x_2}; \mathbf{y_1})}{\partial P_1} = -\frac{\sqrt{snr}h_{11}^2}{(h_{11}^2 P_1 + 1)} \int \nabla_{\mathbf{y_1}} p_{y_1}(\mathbf{y_1}) \times$$
$$\mathbb{E}_{x_2|y_1}\left[\mathbf{x_2}|\mathbf{y_1}\right] h_{12}\sqrt{P_2} d\mathbf{y_1} \quad (64)$$

$$\frac{\partial I(\mathbf{x_2}; \mathbf{y_1})}{\partial P_1} = \int \frac{\sqrt{snr}h_{11}^2}{(h_{11}^2 P_1 + 1)} p_{y_1}(\mathbf{y_1}) \times$$
$$\left(\mathbf{y_1} - \sqrt{snr}h_{12}\sqrt{P_2}\mathbb{E}_{x_2|y}[\mathbf{x_2}|\mathbf{y}]\right) \mathbb{E}_{x_2|y_1}\left[\mathbf{x_2}|\mathbf{y_1}\right] h_{12}\sqrt{P_2} d\mathbf{y_1} \quad (65)$$

$$\frac{\partial I(\mathbf{x_2}; \mathbf{y_1})}{\partial P_1} = \frac{\sqrt{snr}h_{11}^2}{(h_{11}^2 P_1 + 1)} \mathbb{E}_{\mathbf{y_1}}[\mathbf{y_1}\mathbf{x_2}^\dagger h_{12}\sqrt{P_2}]$$
$$- \frac{\sqrt{snr}h_{11}^2}{(h_{11}^2 P_1 + 1)} \sqrt{snr}h_{12} P_2$$
$$\mathbb{E}_{\mathbf{y_1}}[\mathbb{E}_{x_2|y_1}[\mathbf{x_2}|\mathbf{y_1}]\mathbb{E}_{x_2|y_1}[\mathbf{x_2}|\mathbf{y_1}]h_{12}\sqrt{P_2}] \quad (66)$$

Therefore,

$$\frac{\partial I(\mathbf{x_2}; \mathbf{y_1})}{\partial P_1} = \frac{\sqrt{snr}h_{11}^2}{(h_{11}^2 P_1 + 1)} \mathbb{E}_{x_2}[\mathbf{x_2}\mathbf{x_2}^\dagger h_{12} P_2]$$
$$- \frac{\sqrt{snr}h_{11}^2}{(h_{11}^2 P_1 + 1)} \sqrt{snr}h_{12}\sqrt{P_2} \times$$
$$\mathbb{E}_{y_1}[\mathbb{E}_{x_2|y_1}[\mathbf{x_2}|\mathbf{y_1}]\mathbb{E}_{x_2|y_1}[\mathbf{x_2}|\mathbf{y_1}]h_{12}\sqrt{P_2}] \quad (67)$$

$$\frac{\partial I(\mathbf{x_2}; \mathbf{y_1})}{\partial P_1} = \frac{snr h_{11}^2 h_{12}^2 P_2 E_{22}}{(h_{11}^2 P_1 + 1)} \quad (68)$$

Similarly, we can derive the gradient of other terms of the non-conditional mutual information. Therefore, Theorem 3 has been proved.